\documentclass{article}
\usepackage{amsmath}
\usepackage{dsfont}
\usepackage{graphicx}
\usepackage[numbers]{natbib}
\usepackage{xcolor}

\usepackage[fontsize=12pt]{fontsize}
\usepackage
[a4paper, left=3cm, right=3cm, top=3cm, bottom=3cm]{geometry}
\usepackage{wrapfig}
\usepackage{etoolbox}
\usepackage{float}
\usepackage{fancyhdr}
\usepackage{setspace}
\title{Physics-embedded inverse analysis with automatic differentiation for the earth's subsurface}
\author{Hao Wu$^{1,*}$, Sarah Y. Greer$^{1,2}$ and Daniel O'Malley$^{1}$\\
\footnotesize$^1$Computational Earth Science, Los Alamos National Laboratory, Los Alamos, NM 87545\\
\footnotesize$^2$Massachusetts Institute of Technology, Cambridge, MA 02139\\
\footnotesize$^*$corresponding author: wu\_hao@lanl.go}

\begin{document}

\maketitle

\begin{abstract}
Inverse analysis has been utilized to understand unknown underground geological properties by matching the observational data with simulators. 
To overcome the underconstrained nature of inverse problems and achieve good performance, an approach is presented with embedded physics and a technique known as automatic differentiation. 
We use a physics-embedded generative model, which takes statistically simple parameters as input and outputs subsurface properties (e.g., permeability or P-wave velocity), that embeds physical knowledge of the subsurface properties into inverse analysis and improves its performance. 
We tested the application of this approach on four geologic problems: two heterogeneous hydraulic conductivity fields, a hydraulic fracture network, and a seismic inversion for P-wave velocity. 
This physics-embedded inverse analysis approach consistently characterizes these geological problems accurately. 
Furthermore, the excellent performance in matching the observational data demonstrates the reliability of the proposed method. 
Moreover, the application of automatic differentiation makes this an easy and fast approach to inverse analysis when dealing with complicated geological structures. 

\end{abstract}

\section{Introduction}

Inverse analysis gives the solution of inverse problems aiming to find unknown properties of an object, or a medium, from observing a response of this object or medium \cite{ramm2006inverse}. 
The inverse analysis process describes finding the matched predictions through a forward model calculation, which takes the parameters describing unknown properties as input, to the observational data \cite{linde2015,zhou2014}. 
A representative example is seismic inversion, which often involves triggering a source wavefield at the earth's surface and collecting the scattered data at receivers from various positions along the surface. 
Accounting for the received data, it is possible to find the heterogeneous subsurface structures, such as the existence of an oil deposit, a cave, or a mine \cite{ramm2006inverse}. 
In earth science, geological reservoir characterization is of essential value for maximizing oil production from mature hydrocarbon provinces, detecting fluid distributions (groundwater, oil, gas, etc.),  \cite{lake2012reservoir, lake1991review}, and many other important issues affecting our daily lives \cite{carrera1986estimation, sun2013inverse, carrera2005inverse}. 
In addition, reservoir properties show spatial heterogeneities from pore to reservoir scale, and it is critical to exploit well the heterogeneity effects on the underground fluid flow system \cite{doughty2004g, jayne2019geologic}.  
However, since physical properties can not be observed directly in the field, applying inverse analysis techniques to understand the heterogeneous reservoir properties is necessary based on observational data such as pressure for hydraulic conductivity fields. 

In this study, we propose a novel method for inverse analysis, which generalizes different inverse analysis approaches and can include embedded physics understanding. 
In addition, we test this novel inverse analysis method for different earth science problems: heterogeneous hydraulic conductivity of a groundwater flow system, hydraulic fracture distribution in a gas-producing reservoir, and seismic inverse for subsurface aquifer determination. 
Porous media has been the source of valuable fluids such as groundwater and petroleum, as well as both liquid and natural gas \cite{valko1995hydraulic}. 
In a groundwater flow system, fully understanding the heterogeneous subsurface hydraulic conductivity distribution is of importance for estimation of drinking groundwater utilization and contamination mitigation \cite{carrera1986estimation, sun2013inverse, carrera2005inverse}. 
In addition, the earth's subsurface has also been used for the injection of slurried wastes, like hazardous chemicals or radioactive byproducts \cite{pollyea2019high, pollyea2020new}, and certain geological reservoirs have been used for CO\textsubscript{2} storage and recovery \cite{wu2018parametric, jayne2019geologic, jayne2019probabilistic, wu2021multi,wu2021simulation}.
Characterizing the underground structures, which enables the prediction of the fluid flow system behavior, is essential for successfully using geological sources and avoiding environmental contamination for the projects mentioned above. 
Notably, we focus on two different scales of heterogeneity in this study. 
Furthermore, among the producing wells drilled in North America since the 1950s, around 70\% of gas wells and 50\% of oil wells have been hydraulically fractured. 
Once a hydraulic fracture is generated, fluid in the reservoir will flow out or into the fracture face and then, along the fracture path, flow out or into the injection or production well \cite{montgomery2010hydraulic, economides1989reservoir, greer2022comparison}. 
Over the past decades, hydraulic fracture simulation has become a significant part of the design and analysis of oil fields through reservoir characterization and simulation \cite{lecampion2018numerical}. 
This study depicts the hydraulic conductivity distribution of a hydraulic fracture network by applying inverse analysis for future oil production estimation and optimization. 
Last but not least, we exploit a physics-embedded generative model for seismic inversion problems for predictions about underground lithology \cite{russell2006old}. 
Seismic inversion aims to reconstruct the subsurface structure based on seismic measurements, like trapping mechanisms for hydrocarbon reservoirs and fracture distribution for groundwater storage \cite{wang2016seismic}.
During the seismic inversion, reservoir properties of interest, such as lithologies, can be transformed from elastic properties (e.g., velocities), which are inverted from seismic data \cite{bosch2010seismic}.
Here, we bring inverse analysis to crack easy and fast seismic inversion about underground geological layer properties. 

The inverse analysis provides support for underground feature characterization in earth sciences.
Despite its effectiveness, inverse analysis is challenging to conduct, and computationally expensive \cite{carrera2005inverse}. 
In addition, another concern of inverse analysis is that it may lead to many viable solutions resulting in a second-round calibration or investigation due to the underconstrained or ill-posed features \cite{zagst2008inverse}. 
There are many feasible ways to conduct inverse analysis, for example, the geostatistical approach \cite{lee2014large, kitanidis2014principal}, physics-based imaging methods \cite{tarantola1984linearized}, and machine learning (ML) \cite{sinha2020normal, zhou2019data}. 
The stochastic geostatistical inversion approach was recognized at unknown parameters estimation \cite{tartakovsky2021physics}, which describes the unknown properties. 
However, the randomness of the variables reflects the lack of certainty about their values, which are coded as the probability distributions of the quantities. 
As a result, the solution to an uncertainty quantification turns out to be the maximum likelihood probability distribution of the target variables, based on all the information completed to be interpreted \cite{kaipio2006statistical}. 
Given this feature of the traditional geostatistical approach, the high computational cost may be one concern when dealing with large-scale systems. 
Even if some recent developments overcome this challenge, like the principal component geostatistical approach (PCGA) \cite{lee2014large, kitanidis2014principal}, reducing the dimension parameter space by only focusing on principal components of the covariance matrix, the struggle still cannot be avoided when dealing with highly complicated surface structures that are not amenable to a two-point correlation structure. 
In the use of reflection seismic data, various migration methods are often used to map recorded surface data to their corresponding subsurface reflection points. 
More robust methods, such as reverse-time migration and full-waveform inversion, can work on models with complex geologic structures but require significant computational cost \cite{baysal1983reverse, yilmaz2001seismic}. 
An additional representative method is ML, which recently improved inverse analysis in the geological area \cite{barajas2019approximate,kadeethum2021framework}. 
These ML studies did use a physics model during the initial training, which led to a scenario with a steep up-front cost to generate the training data, even if they can produce excellent results \cite{karniadakis2021physics}. 
With this, it has to face some situations where the cost of training data generation is greater than the inverse analysis process, which can be self-defeating in some cases, lying in the limitation that each training data point needs to run through the forward model. 
In addition, the success of the ML models is subject to the appropriate selection of ML structures, while the uncertainty of results cannot be predicted from different structures \cite{mcgovern2019making,karniadakis2021physics}. 
In the work of \cite{aggarwal2018modl}, they applied deep learning for inverse problems about image reconstruction, which illustrates the pressing need for model structure selection, even if there is a considerable improvement in the demand for the training data.

To overcome the aforementioned challenges, including physical calibration during inverse analysis has been successful in enhancing the final results and the model performance \cite{he2020physics, tartakovsky2020physics, geneva2020modeling, zhu2019physics}. 
Take the work of \cite{mohan2020spatio}, for instance; they addressed the physics-informed diagnostics by testing various ML algorithms' capability for turbulence flow. 
Interestingly, this physics-informed ML contributes to taking advantage of the mathematical properties of the underlying physics foundation, yielding interpretable strategies from numerical methods and computational fluid dynamics. 
As a result, it increased the reliability of ML schemes by its high efficiency and accuracy; at the same time, based on putting insights about the understanding of the complicated ML structures \cite{mohan2020embedding}. 
Other similar examples, like the work of \cite{mumpower2022physically, raissi2019physics}, present similar approaches to incorporating knowledge of physics as a soft physics constraint for the loss function penalty in the area of quantum mechanics. 
In earth sciences, physics-embedded inverse analysis has been widely applied.
In our previous work \cite{wu2022inverse}, we proposed an approach to achieve an easy and fast inverse analysis to interpret complex heterogeneous hydrogeologic reservoir properties by applying the variational autoencoder (VAE), which combines the strengths of the traditional geostatistical approach and recent ML techniques.
In addition, we tested the different neural network architectures based on result stability and reliability. 
Similarly, the physics-informed autoencoders have been investigated for underground fluid flow prediction, and it provides a comprehensive understanding of model stability and prediction certainty improvement \cite{erichson2019physics}. 
As a result, involving the understanding of physics foundation by applying the mathematical properties of physics laws when generating the target fields is a powerful tool for inverse analysis problems, which provides fundamental support for accuracy and efficiency enhancement.
Therefore, this study describes our novel approach to physics-embedded inverse analysis and demonstrates its efficacy on multiple subsurface problems, including subsurface flow and seismic wave propagation.

Underground reservoir property characterization is complicated because the underlying system is unknown \cite{trujillo2017practical}. 
For inverse analysis, it is essential to rely on observational data to discover the underground structures and features since it is impossible to directly observe all the detailed information about the whole system in the field. 
Given this reason, especially for large-scale inverse models, the use of many observations is essential \cite{lin2017large}. 
Thanks to the rapid development of sensor networks, we can collect a wealth of variable fidelity observations and monitor the evolution of complex phenomena at large spatial and temporal scales \cite{karniadakis2021physics}. 
Consequently, it leads to a scenario where the inverse analysis based on observational data can be performed. 
Beyond this, another key factor to achieving successful inverse analysis is result calibration by matching observational data. 
Even though enough observational data has been utilized during calibration to reach good performance of the inverse model, regularization, a numerical technique involving adding a term to the objective function, is highly valued at improving results. 
Ideally, a small objective function value indicates good performance. 
Adding a regularization term to the objective function seeks to develop additional desired features to the inverse solution, such as smoothness, convexity, or respecting prior knowledge of geologic features. 
In addition, optimization is the most time-consuming step during inverse analysis, but we apply automatic differentiation to increase the computational efficiency \cite{innes2019differentiable}. 
Automatic differentiation can compute gradients with a low computational cost for complicated computer programs by applying the chain rule repeatedly. 
Specifically, automatic differentiation is good at calculating high-dimensional derivatives, which is often useful for inverse analysis problems with substantial computational savings. 
Because of regularization and automatic differentiation techniques, inverse analysis becomes more feasible to estimate interest quantities reasonably based on available data \cite{kaipio2006statistical}. 

The remainder of this manuscript describes the workflow of the general physics-embedded inverse analysis, the inverse analysis results for different problems, and the benefits and improvement of this approach in sections 2, 3, and 4. Finally, we present our conclusion about applying this approach in section 5. 

\section{Methods}\label{sec:methods}

The physics-embedded inverse analysis starts with the physics-embedded generative model generation. 
Specifically, in this study, the physics-embedded generative models describing the quantities of interest are the heterogeneous hydraulic conductivity distribution, hydraulic fracture distribution, and seismic P-wave velocity. 
These are the targets of what inverse analysis is trying to predict through observational data matching.
Several key factors are picked to represent the variability of the targets through the physics-embedded generative model for the stability test. 
For example, in the hydraulic fracture problem, five key factors (which can be understood as latent variables) are utilized to represent the lengths of the hydraulic fractures in a cluster, which is correlated with permeability. 
The physics-embedded generative model describes the relationship between the latent variables to the target properties we are interested in. 
In addition, the physics-embedded generative model embeds physical knowledge of the system. 
Continuing with the hydraulic fracture example, once the lengths of the hydraulic fractures are figured out, the permeability of the fractures could be calculated based on the mathematical models, e.g., we apply the fracture size-transmissivity relationship. 
Finally, we build a model that characterizes the hydraulic fracture permeability distribution based on the representative latent variables. 
Broadly speaking, the physics-embedded generative model links the small number of latent variables to many target properties of interest, encoding the relationship between them. 
The physical knowledge embedded in the generative model increases the reliability and accuracy of the inverse analysis.  

\begin{figure}
	\includegraphics[width=\textwidth]{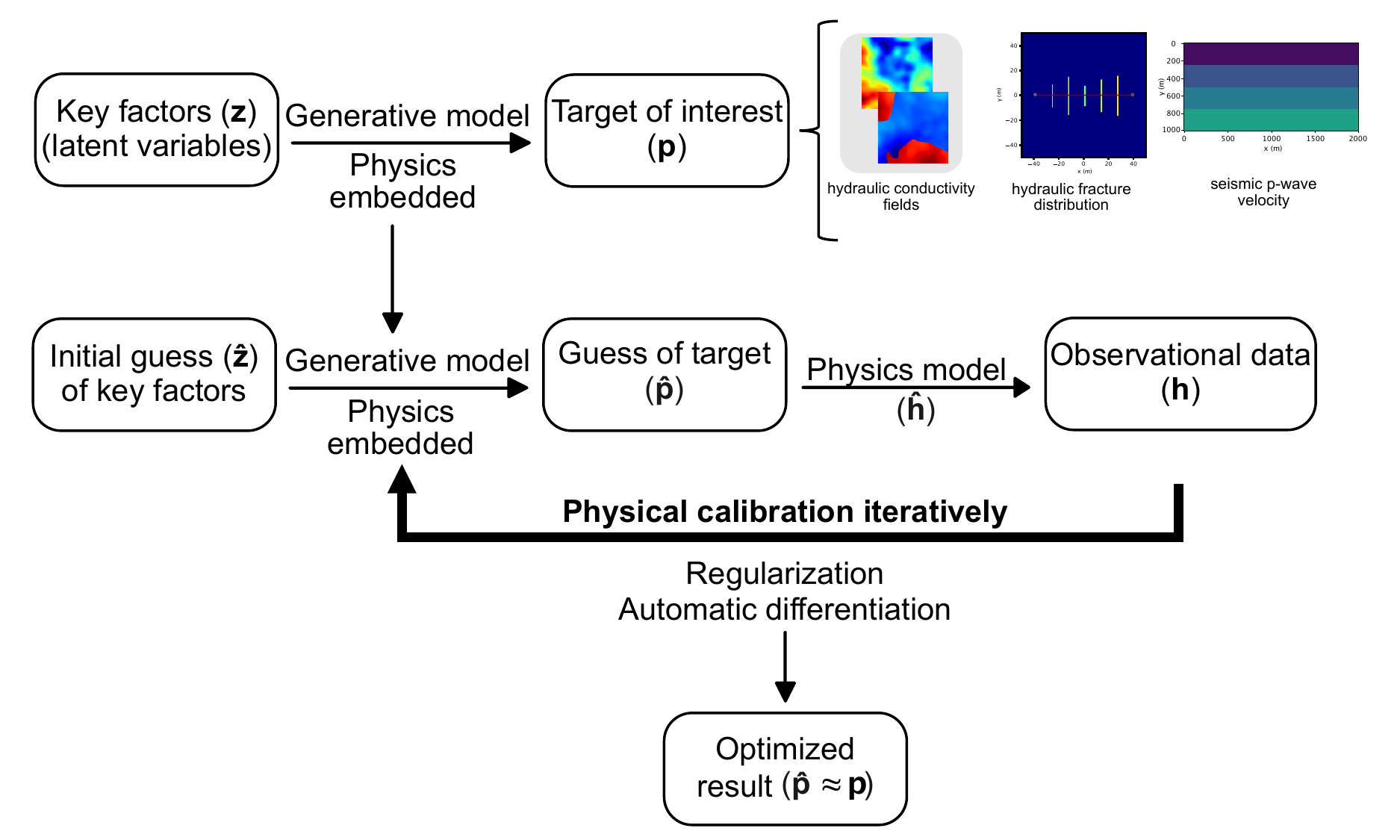}
	\caption{physics-embedded inverse analysis leverages the underground geological property characterization with physics law understanding and automatic differentiation. Inverse properties are calculated based on the physics-embedded generative model, and calibration is performed on the output of the forward model. The application of regularization and automatic differentiation makes the optimization process simple.}
	\label{fig:workflow}
\end{figure}

Once the physics-embedded generative model has been constructed, the second step is the objective function set up, a forward physical model taking the output from the physics-embedded generative model to simulate the mechanisms of the study system, like fluid flow for hydraulic conductivity fields.
Using the hydraulic fracture problem as an example again, a gas production situation has been simulated.
The fluid flow from fractures to the pumping well, specifically the pressure drop, is calculated through the forward model. 
The objective function characterizes the difference between the observational data and the predicted output from the forward model. 
The final step is performing the inverse analysis using gradient-based optimization with the gradients being computed automatically.
During this step, the output from the forward model is compared with the observational data through the loss function to achieve final optimized results.
The detailed workflow is illustrated in figure \ref{fig:workflow}.

During this study, the notation $\mathbf{p}$ and $\hat{\textbf{p}}$ are used to represent a physical reservoir property field in vector form, $\mathbf{z}$ and $\hat{\textbf{z}}$ represent the latent variables, and $\mathbf{h}$ and $\hat{\textbf{h}}$ represent a vector of observations for inverse analysis and the calculation from the forward physical model, respectively. 
This study obtained observations $\mathbf{h}$ from different problems directly from the related reference fields $\mathbf{p}$ through the forward model. 
The forward physical model predictions $\hat{\textbf{h}}$ are obtained based on the guess of target properties $\hat{\textbf{p}}$ through the iterations.  
The error used to measure the difference between the true and predicted values should be differentiable for the inverse analysis. 
This loss takes the simplified form of the sum of squared residuals in the examples studied here. 

The optimization problem of inverse analysis is formulated in terms of the key factors or latent variables, $\hat{\textbf{z}}$ and includes regularization in the objective function. 
Regularization implemented additional desired features to the final solution, avoiding side effects like overfitting. 
In addition, automatic differentiation is utilized to compute the objective function, resulting from its ability to efficiently calculate the gradient. 
Specifically, the automatic differentiation library we applied is Zygote.jl \cite{innes2019differentiable}, and we use the differentiable physics simulator, DPFEHM \cite{DPFEHM}. 
As for optimizing the objective function, a gradient-based optimization method is utilized, which is the limited-memory Broyden–Fletcher–Goldfarb–Shanno \cite{liu1989limited} (L-BFGS) method with a Hager–Zhang line search \cite{hager2005new}. 
The Optim.jl \cite{mogensen2018optim} software package is specified for this process.
Of course, other gradient-based optimization routines could also be used.
For all the problems, to start the inverse analysis, the initial guess of key factors is set to be 0.

\section{Examples and Results}\label{sec:results}

This study provides a generative approach to physics-embedded inverse analysis. 
We focus on three problems: heterogeneous hydraulic conductivity field, hydraulic fracture distribution, and seismic inversion of P-wave velocity property. 
Two types of heterogeneity have been considered for heterogeneous hydraulic conductivity fields. 
For larger-scale heterogeneous fields, to improve the inverse analysis performance, the ML method was applied. 
To estimate the performance of inverse analysis, the comparison of reference fields and the final inverse results are conducted for different problems are presented in figures \ref{fig:PCGA_fig}, \ref{fig:bimodal_fig} - \ref{fig:wave_fig}. 
Figures \ref{fig:PCGA_head}, \ref{fig:RegAE_head} - \ref{fig:wave_wave} show the comparison of observational data and the prediction of the forward physical model after inverse analysis. The convergence of different inverse analyses showing the optimization process is described in the supplementary information. 

In figure \ref{fig:bimodal_fig} - \ref{fig:wave_fig}, the comparison results, the first rows are the three ``true'' reference fields. 
The following rows demonstrate the inverse results, while the last rows are the difference calculated between the ``true'' and results estimated by the inverse analysis. 
On top of each simulated final result is the relative error, which measures how close the inverse result is to the reference field and is defined as 
\begin{equation}
	\frac{||\mathbf{p} - \hat{\mathbf{p}}||^2}{||\mathbf{p} - \bar{\mathbf{p}}||^2}\;, 
    \label{eq:error}
\end{equation}
where $\bar{\mathbf{p}}$ is the mean of the reference field. 
Especially for the Gaussian hydraulic conductivity field in figure \ref{fig:PCGA_fig}, only one reference field has been represented to investigate the performance of the proposed approach. 

\subsection{Principal component geostatistical approach for Gaussian hydraulic conductivity}

\begin{figure}
	\includegraphics[width=\textwidth]{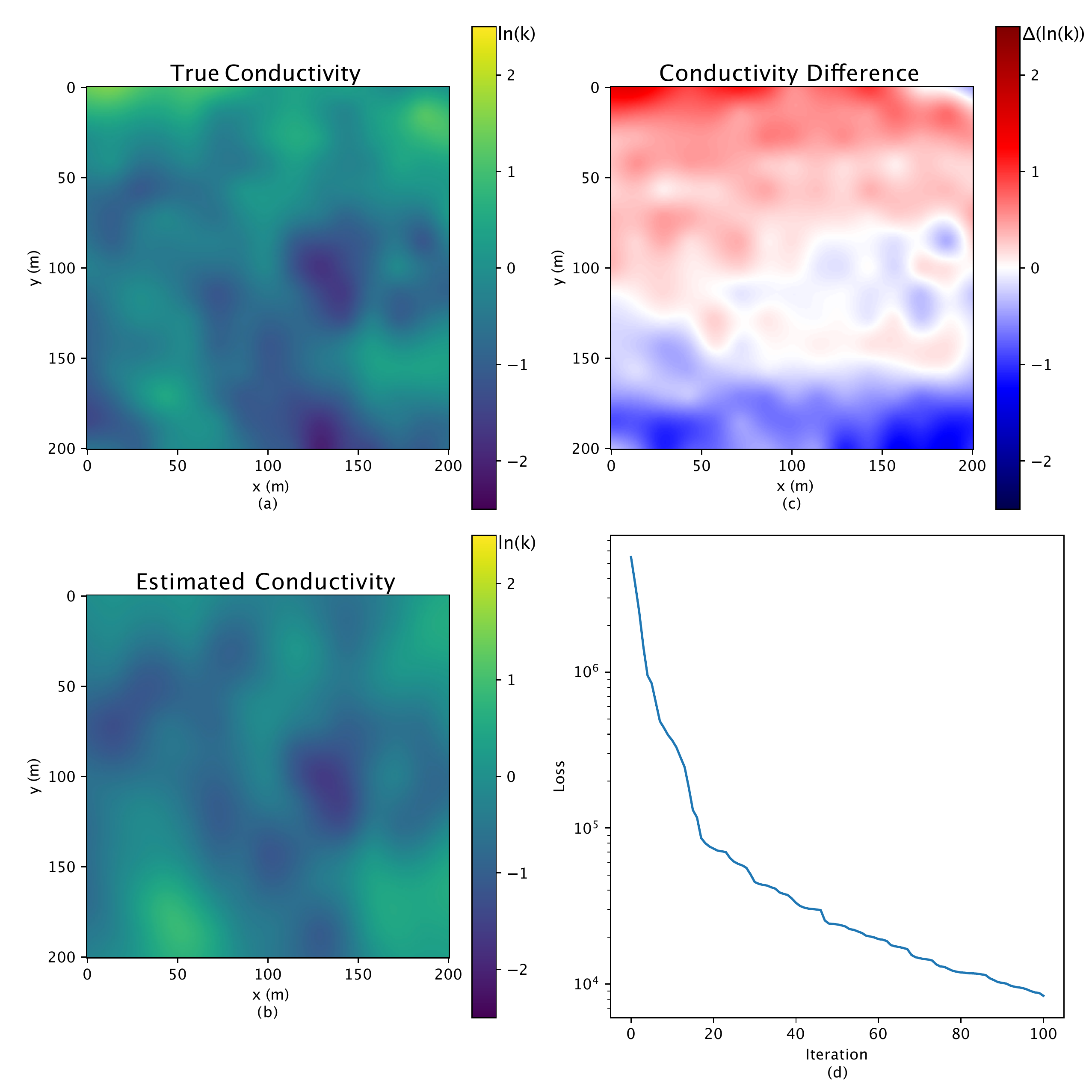}
	\caption{The reference conductivity field is shown in subfigure (a), and the corresponding inverse results are shown in subfigure (b). The hydraulic conductivity shows on the $ln$ scale. Subfigure (c) shows the difference between the related reference fields and the inverse results. The convergence of the inverse analysis is shown in subfigure (d).}
	\label{fig:PCGA_fig}
\end{figure}

One of this work's focuses is a heterogeneous hydraulic conductivity field. 
First, we discuss a multivariate Gaussian field of small-scale heterogeneity in the hydraulic conductivity field. 
A 200 m $\times$ 200 m subsurface aquifer is simulated with a unit thickness.
Two hundred eigenvalues $\mathbf{z}$ (the principal components) have been introduced to go through a Gaussian distribution, with mean 0, variance 1, and correlation length 50 m, to create the heterogeneity of the research area. 
We use the GaussianRandomFields.jl package to generate the multivariate Gaussian field for the Julia programming language\cite{GaussianRandomFields}. 
The heterogeneous hydraulic conductivity field $\mathbf{p}$ is shown in figure \ref{fig:PCGA_fig} (a). 
For the physics-embedded generative model, we utilize the principle components of the covariance matrix to represent the Gaussian distribution.
This embeds knowledge of the statistical structure of the permeability fields.

\begin{figure}
	\includegraphics[width=\textwidth]{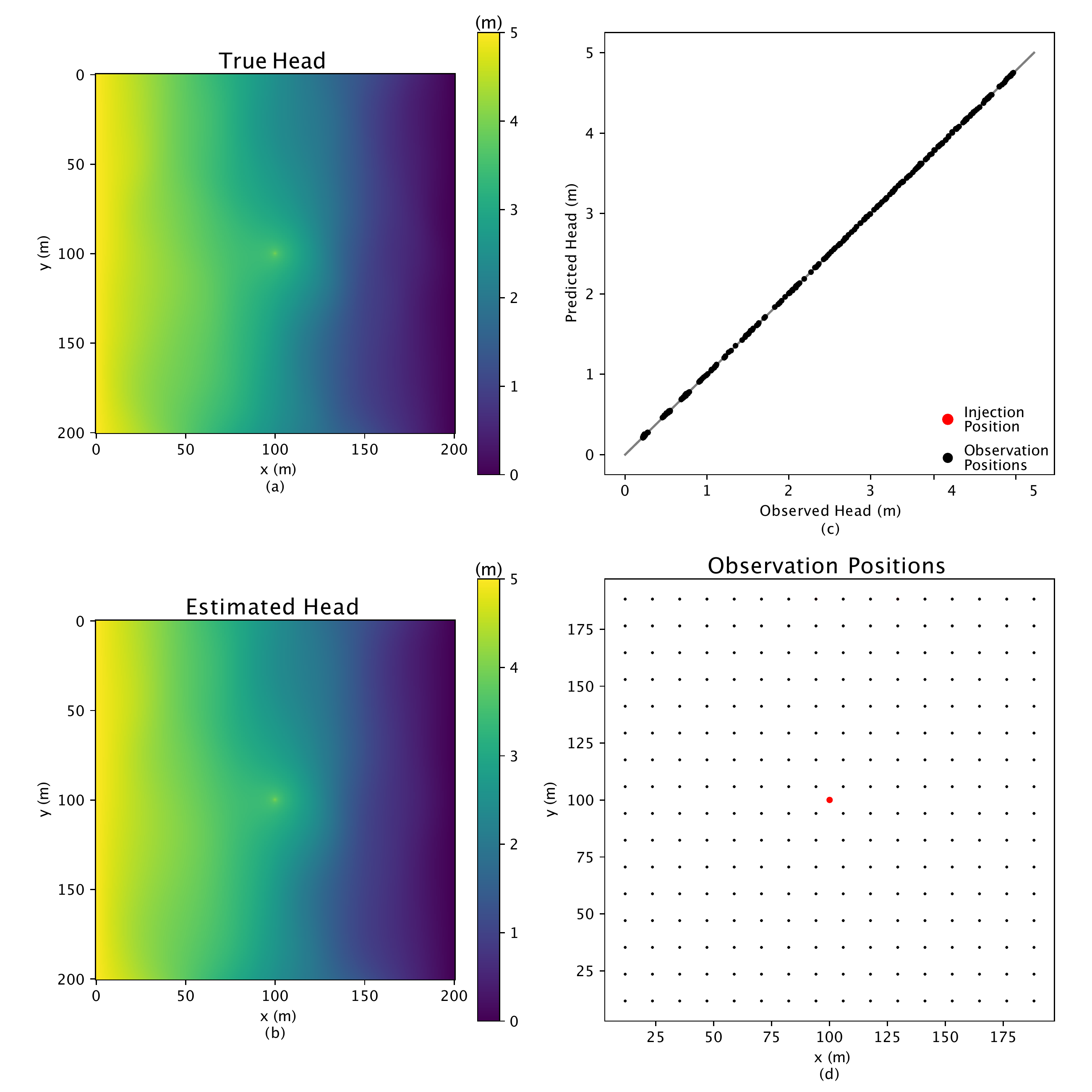}
	\caption{The head distribution of the reference conductivity field is shown in subfigure (a), and subfigure (b) shows the corresponding inverse estimated head distribution. The black dots in subfigure (d) represent the 16 × 16 regular grids for observation points, and the one red dot illustrates the injection position, with a rate of 1.0 $m^3$/s. Finally, the comparison of the observed head and predicted head for all the observational points are shown in subfigure (c).}
	\label{fig:PCGA_head}
\end{figure}

More importantly, the background groundwater flow is simulated through a forward physical model for inverse analysis calibration. 
The boundary condition yields a constant 5 m head drop from left to right. 
In addition, in the center of the research area, water is injected at a rate of 1.0 $m^3$/s.
The observation used to inform the inverse analysis is the hydraulic head, from a static forward Darcy's law, on a 16 $\times$ 16 regular grid within the domain. 
Figure \ref{fig:PCGA_head} (a) shows the reference head distribution, and the positions for all the observations are shown in figure \ref{fig:PCGA_head} (d). 

For the Gaussian hydraulic conductivity field in figure \ref{fig:PCGA_fig},  the ``true” reference field and the estimated result are similar, particularly for the center area.
The subfigure \ref{fig:PCGA_fig} (c) shows their difference, which is mainly in the center and relatively small, and indicates the good performance of the inverse analysis. 
However, the error reaches around one order of magnitude at the top and bottom edges. 
This is partly due to the fixed pressure boundary conditions, which make the observations less sensitive to the hydraulic conductivity near these boundaries.

\subsection{ML approach for bimodal hydraulic conductivity}

\begin{figure}
	\includegraphics[width=\textwidth]{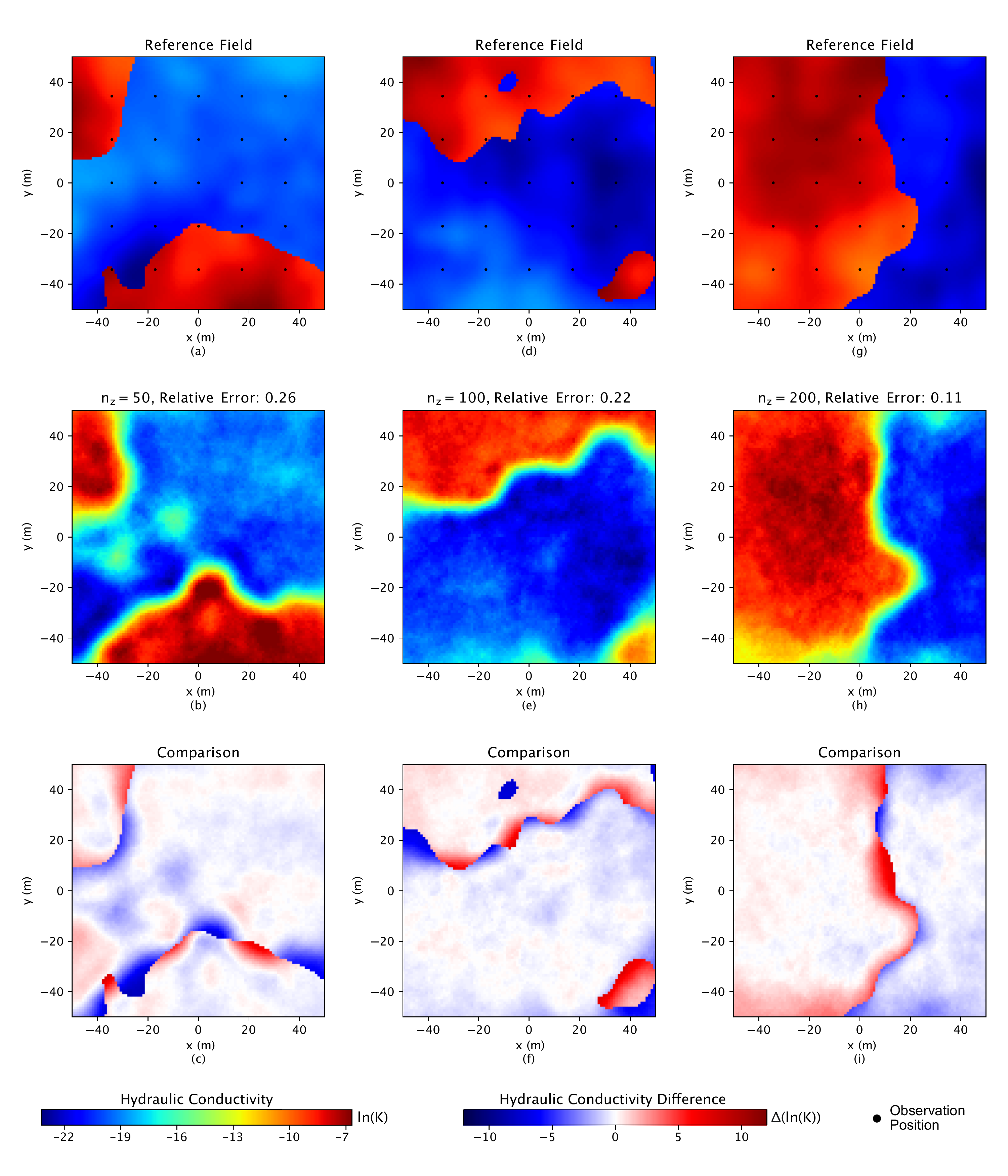}
	\caption{Three bimodal reference conductivity fields are shown in subfigures (a), (d), and (g). The black dots in the first row represent the 5 × 5 regular grids for observation points. The corresponding inverse results are shown in the second row, subfigures (b), (e), and (h). The hydraulic conductivity shows on the $ln$ scale. The relative errors above inverse figures represent the difference between related reference and simulation figures. The last row shows the difference between the related reference fields and the inverse results.}
	\label{fig:bimodal_fig}
\end{figure}

Beyond the Gaussian field, to show our approach generalizes additional methods, we show how it generalizes RegAE. 
RegAE is a method that can solve more challenging permeability fields than the principal component geostatistical approach. 
The domain is a 100 m $\times$ 100 m subsurface aquifer with a unit thickness. 
In this type of field, the higher heterogeneity is applied and is represented by two hydrogeologic facies with distinct properties, each of which is a multivariate Gaussian distribution. 
The two multivariate Gaussian structures are shown as conductivity 1 and 2 in table \ref{tab:fieldparams}.
More importantly, the ``Split'' model has a different multivariate Gaussian structure that has been utilized to indicate which of the facies is present at a given location.  
As a result, the new type of field shows a bimodal hydraulic conductivity distribution, and the reference fields are represented in figure \ref{fig:bimodal_fig} (a), (d), and (g). 

\begin{table}[]
	\begin{tabular}{|l|l|l|l|}
		\hline
		Model Name & Mean $(m/s)$ & Variance $(m^2/s^2)$ & Correlation length (\textbf{m})\\
		\hline
		Conductivity 1 & $10^{-5}$ & 1.0 & 50.0 \\
		Conductivity 2 & $10^{-8}$ & 1.0 & 50.0\\
		Split & $10^{-8}$ & 1.0 & 200.0 \\
		\hline
	\end{tabular}
	\caption{The statistical parameters used to generate the hydraulic conductivity data.}
	\label{tab:fieldparams}
\end{table}

In terms of the higher heterogeneity of the bimodal fields, a generative machine learning model VAE is included to capture the hydrogeological properties distribution in this study. 
VAE \cite{kingma2013auto} is a generative ML model with neural network architecture and has widespread application for image data \cite{doersch2016}. 
VAE consists of two parts: encoder and decoder. 
The encoder step maps a high-dimensional space $\mathbf{p}$ (such as pixels in an image) into a smaller parameter space. 
Specifically, the smaller parameter space is the key factor containing the features of the image, which is the hydraulic conductivity distribution in this study, and used to be called the latent variables $\mathbf{z}$. 
In reverse, the decoder maps the latent variables back to their original high-dimensional space. 
For the bimodal situation, the reference fields have a resolution of pixels of 100 $\times$ 100, which is a high dimensional space.
After the training of VAE, $\mathbf{z}$ is applied to represent the physics meaning of the hydraulic conductivity distribution feature. 
At the same time, the decoder step accounts for the physics-embedded generative model. 
Here, the physics is embedded through the process of training the VAE on images that contain the physical understanding of the subsurface---in this case, the two facies.
In this study, three $\mathbf{z}$ values, 50, 100, and 200, are tested for the inverse analysis performance in terms of key factors. 

Like the Gaussian case, for bimodal fields, a constant head drop of 1 m from the left boundary to the right has been set up for the fluid flow system. 
The head distribution is calculated through the forward physical model of groundwater dynamics at a static state. 
Oppositely, a coarse 5 $\times$ 5 regular grid within the domain is demonstrated for observation during inverse analysis calibration, shown in \ref{fig:bimodal_fig} (a), since VAE is a powerful tool to capture the property distribution with less input information. 
A more detailed description of inverse analysis for Gaussian and bimodal fields is specified in \cite{wu2022inverse}. 

More heterogeneous, bimodal hydraulic conductivity fields are simulated through VAE as the physics-embedded generative model, which is good at spacial feature characterization, to test our idea of the physics-embedded inverse analysis; the results are shown in figure \ref{fig:bimodal_fig}. 
For bimodal fields in figure \ref{fig:bimodal_fig}, the broad similarity in each facies between the reference fields and simulated results implies that the inverse analysis approach captures the salient aspects of the hydraulic conductivity distribution features.
Even if the relative error is comparably higher than those from Gaussian fields, the phenomena lie in the higher complexity of bimodal fields.
Meanwhile, for the difference in subfigures \ref{fig:bimodal_fig} (c), (f), and (i), the major error only occurs at the edge of the two facies; oppositely, in each face, the difference is small and close to zero. 
In conclusion, including the physics-embedded generative model during inverse analysis, even for the ML-specific approach, turns out to be a good application for different types of heterogeneous hydraulic conductivity fields based on having consistently good results. 
However, even if it approves the application of the physics-embedded inverse analysis, specifically for more complicated fields, like the edges in bimodal fields, it needs more calibration or on-site investigation for future field applications.  
On the other hand, it also implies that more detailed physics understanding or background should be included when dealing with complicated field situations. 

\subsection{Hydraulic fracture network}

Most drilled wells have been fractured in the oil and gas production field, resulting from fluid pressure differences \cite{montgomery2010hydraulic,valko1995hydraulic,economides1989reservoir}. 
Fully understanding the distribution and properties of the hydraulic fractures is of essential importance for production estimation and reservoir protection. 
In the cases of drilled wells, they are now turned fully horizontally into the target geologic formations.
At the same time, for almost all depths of interest, the hydraulic fracture will be normal to the direction of the horizontal well. 
In this study, a cluster with five hydraulic fractures has been selected to present the process of inverse analysis during gas production \cite{gunaydin2021laboratory}.
A medium-scale matrix of size 100 m $\times$ 100 m, with a 78 m drilled well in the center position, was used to represent the research domain. 
Hydraulic fractures are in the normal direction to the drilled well and distributed in a constant interval between them. 
The length of the hydraulic fractures follows a power law, given by \cite{o2016does} 

\begin{equation}
	f(r) = {\frac{(1-p)}{(\mathbf{R}_1)^{(1-p)}-(\mathbf{R}_0)^{(1-p)}}}*r^{(-p)}\;,
    \label{eq:possibility}
\end{equation}

where $r$ is the length of hydraulic fracture, $p$ is the power, $\mathbf{R}_1$ and $\mathbf{R}_0$ are the maximum and minimum of the hydraulic fracture length range, and $f(r)$ is the possibility of a certain length $r$. 
Based on a literature review of fracture length distributions \cite{bonnet2001scaling}, the power $p$ is set up to be 1.8, while the length range of fractures spans from 10 m to 90 m in this problem.
After the fracture length has been determined, a size-transmissivity relationship, which describes the transmissivity of fractures and shows a positively correlated power law with the length of fractures, is introduced. 
The size-transmissivity relationship is defined in \cite{dershowitz2003aspo, hyman2016fracture} as 

\begin{equation}
	log(T) = log(\alpha * r^{\beta})\;,
    \label{eq:transmissivity}
\end{equation}

where $T$ is the fracture transmissivity, and $\alpha$ and $\beta$ are related parameters with values $1.3*10^{-9}$, and 0.5, respectively. 
In this study, the reservoir has a 10 m thickness. 
In addition, the permeability of the matrix and the drilled well are set to be at the scale of $10^{-22}$ $m^2$ and $10^{-10}$ $m^2$, respectively. 
Specifically, at the two tips of the fractures, a harmonic mean is introduced to represent the change from fracture to the matrix. 
Hence, five random key factors were selected to represent  the possibility of five fracture lengths.
Then the permeability of the fractures in the cluster is calculated through the physics-embedded generative model mentioned above for the following inverse analysis. 

\begin{figure}
	\includegraphics[width=\textwidth]{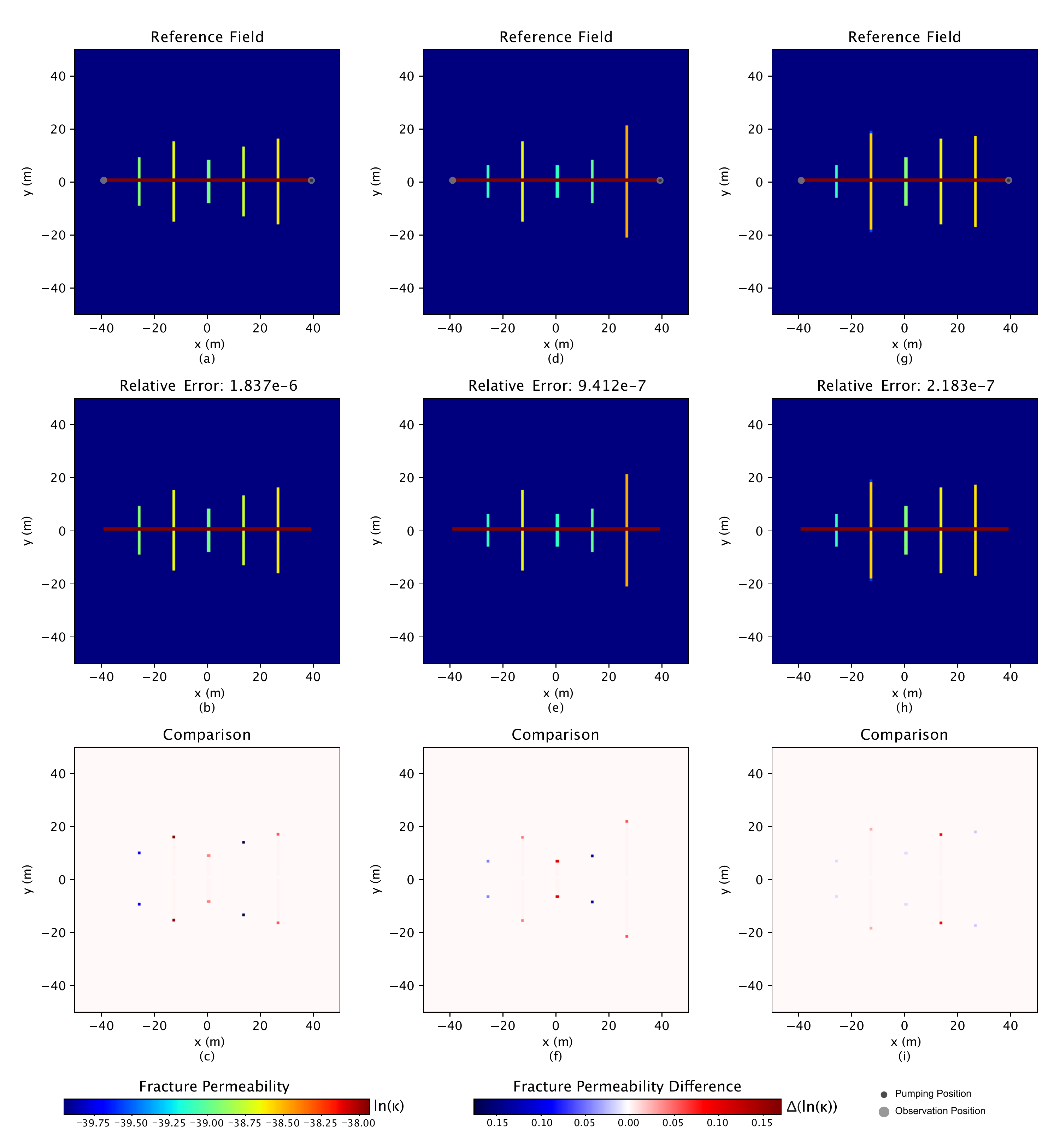}
	\caption{Three hydraulic fracture reference conductivity fields are shown in subfigures (a), (d), and (g). The two grey dots in the first row represent the positions for observation, and the black dot illustrates the gas pumping position. The corresponding inverse results are shown in the second row, subfigures (b), (e), and (h). The permeability shows on the $ln$ scale. The relative errors above inverse figures represent the difference between related reference and simulation figures. The last row shows the difference between the related reference fields and the inverse results. The color bars only apply for the hydraulic fractures, and the permeability for the matrix and well are -52.16 and -24.53 ($ln$ scale), respectively. }
	\label{fig:fracture_fig}
\end{figure}

In this study, gas production has been simulated. 
Especially the gas pumping position is set to be at the right end of the drilled well with a rate of 0.82 $m^3/s$ based on data from the Marcellus Shale Energy and Environment Laboratory (MSEEL) \cite{MSEEL}.
The shale layer of gas production is located at a depth of 2300 m, and the temperature and pressure of the subsurface are 75 $^{\circ}$C and 15 MPa, respectively. 
A transient flow model based on mass conversation and Darcy's law is built for the pressure drop calculation, assuming single-phase gas flow based on the experience at MSEEL, which is very dry. 
Two observation points are located at the two ends of the drilled well, as shown in figure \ref{fig:fracture_fig}. 
As the pumping goes on, the observation lasts for two weeks, with a 30-minute frequency of data collection. 

Similar results in the Gaussian fields for the hydraulic fracture problem are shown in figure \ref{fig:fracture_fig}. 
Surprisingly, the high similarity between the reference fields and the simulated results illustrates the characterizing ability of inverse analysis when including the physics understanding of the relationship between fracture distribution and permeability. 
In addition, the extremely low relative error additionally supports the conclusion. 
The error in the permeability of the hydraulic fractures mainly exists at the two ends of the fractures and is small. 
The more we understand the physics background in the hydraulic fracture problem, the higher possibility we can predict the underground fluid flow system and make a more reliable estimation of oil and gas production.

\subsection{Seismic inversion}

\begin{figure}
	\includegraphics[width=\textwidth]{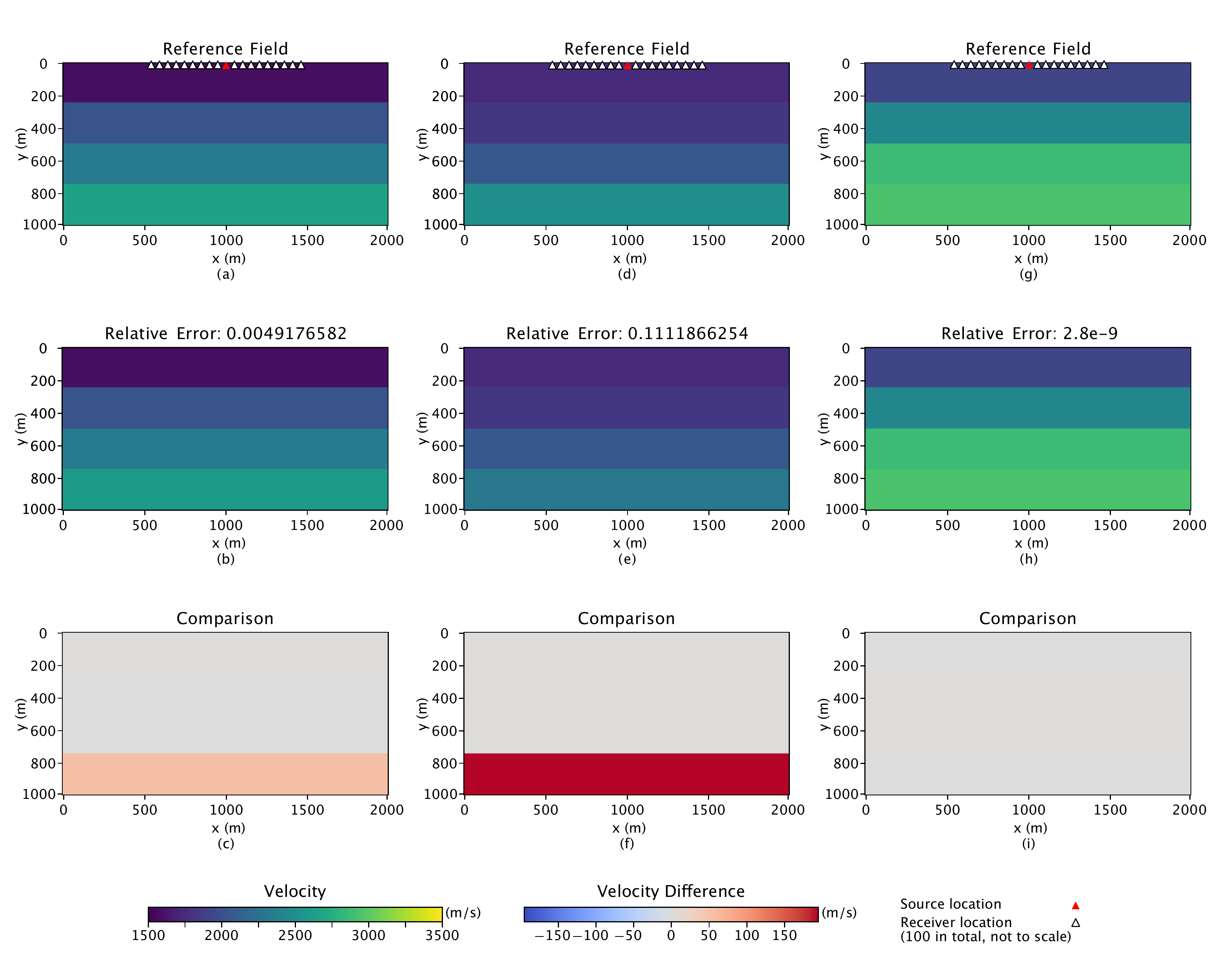}
	\caption{Three reference seismic P-wave velocity fields are shown in subfigures (a), (d), and (g). The red triangle in the first row represents the positions for source location, and the white triangles illustrate the 100 receiver locations, which are not to the real scale. The corresponding inverse results are shown in the second row, subfigures (b), (e), and (h). The relative errors above inverse figures represent the difference between related reference and simulation figures. The last row shows the difference between the related reference fields and the inverse results.}
	\label{fig:wave_fig}
\end{figure}

Seismic inversion estimates subsurface properties by matching predicted data generated on a proposed model to observed data collected at receiver locations.
This study only tests the inverse analysis from the seismic records $\mathbf{h}$ observed at the surface to elastic properties $\mathbf{p}$, which is the velocity at which the P-wave passes through subsurface layers.
The research area is an underground reservoir of size 2 km $\times$ 1 km, consisting of four horizontal layers.
In addition, our physics-embedded generative model embeds the domain knowledge that the velocity tends to increase with depth. 
Four key factors are developed to compute the increasing velocity trend with depth.
The reference fields of the geological layer properties are shown in figures \ref{fig:wave_fig} (a), (d), and (g). 

To generate the observed data, a seismic wave has to be triggered; in this application, the location of the source point is fixed at the center of the domain on the surface. 
Since we constrain our models to only vary with depth and not horizontally, we only use one source per model in our experiments.
In this study, the wave has been computed using a finite difference model.
In addition, 100 seismic receivers are located symmetrically beside the source position along the surface, as shown in figure \ref{fig:wave_fig}. 
Data are recorded at the receivers for a record length of 0.8 s. 
The data from the 100 receivers for the whole simulation time has been implemented for calibration during inverse analysis. 

Finally, we also investigate seismic inversion by applying the physics-embedded inverse analysis. 
Not surprisingly, the simulated data from the estimated model closely matches the reference field data, indicating a good match between the reference and estimated models.
The maximum error is around 11\%, which is acceptable for the deep layers. 
It is more convincing when considering the difference subfigures; the only variance shows for the deepest layer, while all the shallow layers illustrate outstanding results. 
Hence, we can conclude that the generative approach of physics-embedded inverse analysis is successful for the seismic inversion problem. 
Meanwhile, one conclusion from these results is that shallow layers have higher reliability during the inverse analysis, and more investigation or calibration is needed when facing deep layers. 

\begin{figure}
	\includegraphics[width=\textwidth]{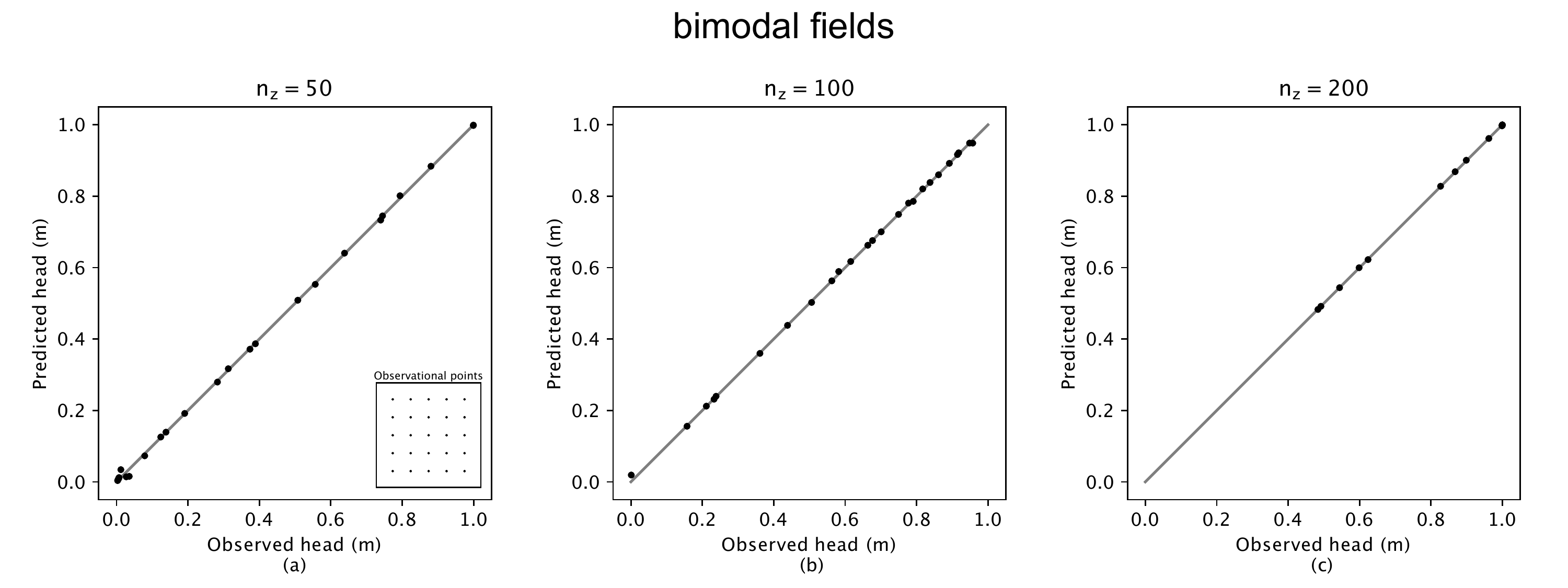}
	\caption{The observed head and predicted head were compared for all the
    observational points. The results illustrate the comparison for bimodal fields. The positions of the observational points are shown in the bottom right corner of subfigure (a). For each field, three amounts of latent variables $\mathbf{z}$ are tested: $n_z$ = 50, 100, and 200.}
	\label{fig:RegAE_head}
\end{figure}

\begin{figure}
	\includegraphics[width=\textwidth]{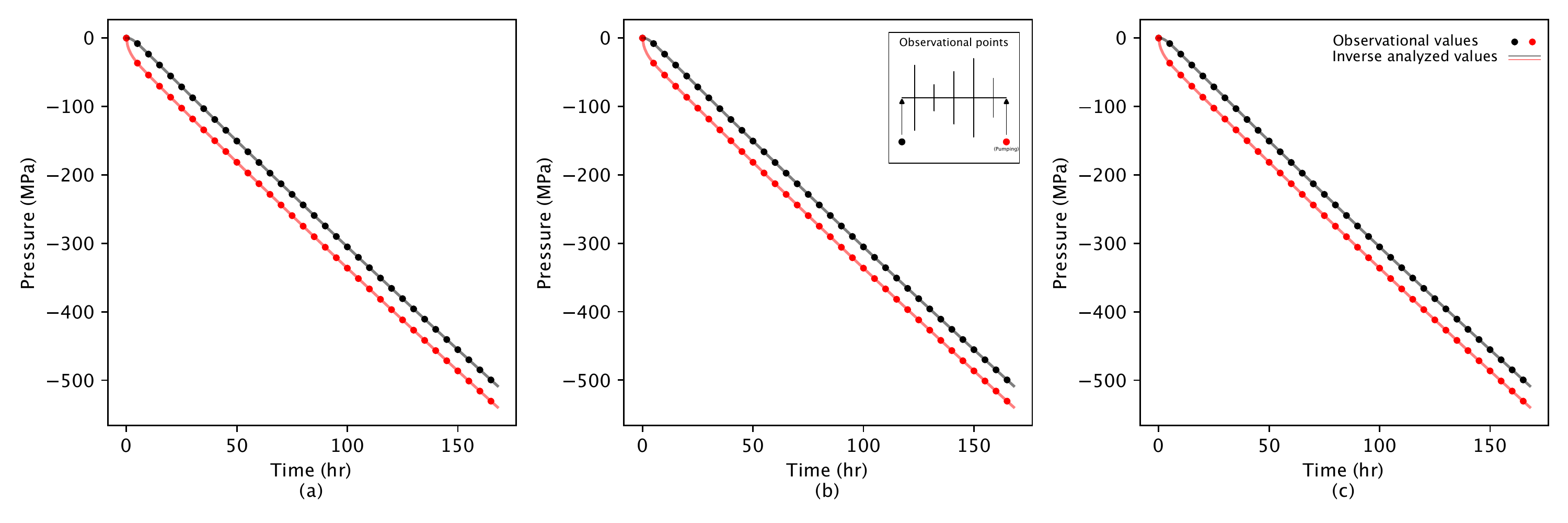}
	\caption{The observed and predicted pressure drops were compared for the two observational points. The positions of the observational points are
    shown in the top right corner of subfigure (b). The dots represent the observational data, and the solid lines are for the inverse results from the forward model. The black dots and solid black line represent the left observational point, while the red ones are for the right. The observation time lasts for two weeks.}
	\label{fig:fracture_head}
\end{figure}

\begin{figure}
	\includegraphics[width=\textwidth]{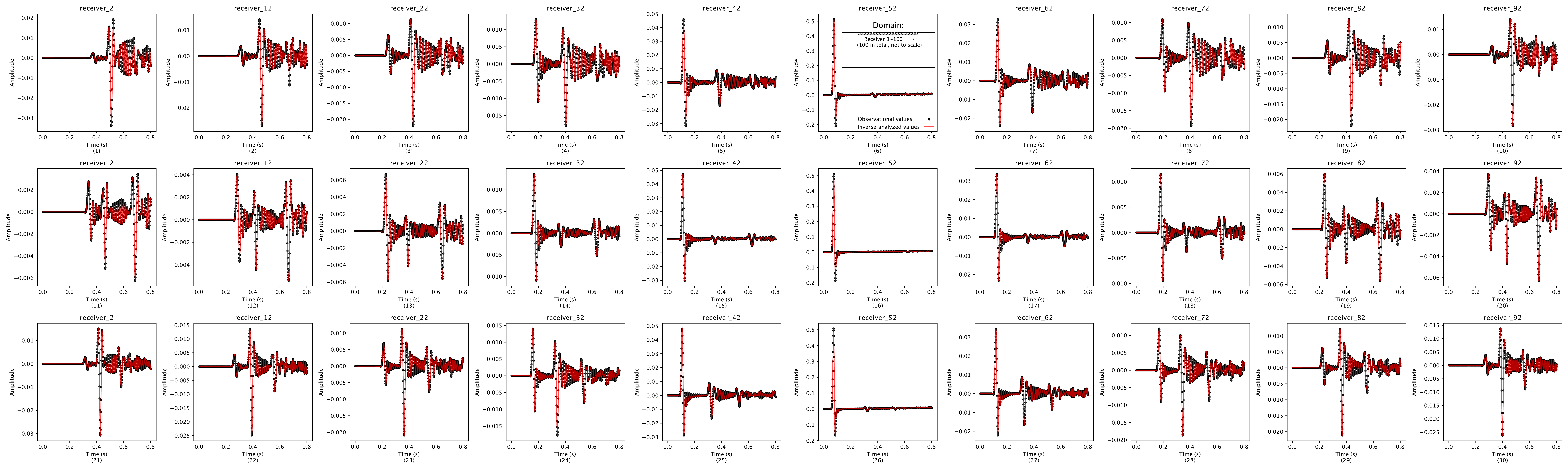}
	\caption{Comparing the observed seismic data and predicted seismic data
    for ten receivers. The positions of the receivers are shown in the top right corner of subfigure (6). There are 100 receivers and lines from 1 to 100 left to right. Only ten receivers are selected to represent the results. The selected ten receivers are from No. 2 and every ten increments (12, 22, and so on). The black dots represent the observational data, and the red line is for the inverse results from
    the forward model. The observation time lasts for 0.8 s.}
	\label{fig:wave_wave}
\end{figure}

\subsection{Comparison of observational results}

Beyond only the comparison between the ``true'' reference fields and the simulated inverse results, the comparison between the observational data and outputs of the forward model is represented in figures \ref{fig:PCGA_head}, \ref{fig:RegAE_head} - \ref{fig:wave_wave} to further demonstrate the performance of the physics-embedded inverse analysis. 
Figures \ref{fig:PCGA_head} and \ref{fig:RegAE_head} compare the observed head and predicted head for all the observational positions, which are shown in the corresponding figures. 
Meanwhile, the hydraulic fracture problem in figure \ref{fig:fracture_head} shows the comparison of observational data and the inverse data about the pressure drop at the observational locations for the two-week simulation time. 
Finally, in figure \ref{fig:wave_wave}, ten receivers are selected to compare the observational and inverse data for the simulation time of 0.8 s. 
The locations of the ten picked receivers are depicted in the figure.
After reviewing all the observation data comparison figures, an inevitable conclusion is that the inverse analysis successfully captures all the observational information features. 
Therefore, it further approves the application of the physics-embedded inverse analysis for underground reservoir property characterization based on observational information on the earth's surface. 

\subsection{Convergence}

The convergence results are depicted in the supplementary information in figures \ref{fig:PCGA_head} and S1 - S4. 
The convergence is generally obtained from 20 to 120 iterations for all the problems.
Generally, the higher accuracy (hydraulic fracture problem) needs more iterations to reach good results; however, this relationship also is affected by the complication of the reference fields and the quality of the physics-embedded generative model.
More discussion about the computational time and cost is in the section Discussion.

\section{Discussion}\label{sec:discussion}

The physics-embedded inverse analysis provides an approach including the physics background understanding to perform inverse analysis effectively. 
At the same time, the application of automatic differentiation shows fast and efficient gradient calculation during optimization.
Our goal here is to demonstrate an inverse analysis approach that uses a physics-embedded generative model by showing how it generalizes some existing methods and can be used more broadly in both subsurface flow problems and seismic inverse problems. 
We now focus our discussion on results and benefits that offer future application potential.
Three inverse problems are completed to investigate the accuracy of the proposed approach; mainly, for the heterogeneous hydraulic conductivity fields, we discussed the two types of heterogeneity and the various physics-embedded generative models during the inverse analysis. 
Generally, the comparison results in figure \ref{fig:PCGA_fig}, \ref{fig:bimodal_fig} - \ref{fig:wave_fig}, which show the high similarity between the reference fields and the simulated results, provide convincingly support for discovering the underground properties using these inverse methods. 
However, for Gaussian hydraulic conductivity problem, which is based on the statistical characterization of the system as the physics understanding, brings some concerns about accuracy only at the edges of the research area. 
At the same time, for the more heterogeneous problem - the bimodal fields, including the VAE method, which can thoroughly characterize the property distribution in each face from image data, illustrate its strong capability for the complicated scenario. 
However, the VAE leading inverse analysis struggles at the boundary of the two facies. 
Therefore, it shows that more complicated problems need a complete understanding of the physics background to reach the perfect performance of the inverse analysis.

Specifically, the hydraulic fracture problem provides the best inversion results of all the research problems. 
Understanding the physics mechanism of the fractures generation and distribution convinced us of its immense potential for highly successful inversion of the hydraulic fracture network, which provides the following estimation or protection plans for oil and gas production.
However, in this study, the two ends of the fracture also draw attention to more calibration. 
It indicates the difficulty of inversion when considering the connection of the hydraulic fracture network with the existing natural fracture system. 
In addition, we only pick one cluster of hydraulic fractures to conduct the inverse analysis; in an actual situation, the production well is several km long, where there are many hydraulic fracture clusters along with it. 
Similarly, in the scenario about seismic inversion, the low relative error and high similarity illustrate the success of the physics-embedded inverse analysis approach.
The error mainly focuses on the deep layers, even if the error is relatively small, which indicates that more consideration may be needed for deeper
Again, there are large fractures, caves, and mines in the subsurface environment, which creates the discontinuity of the properties of interest and brings difficulty to the inversion. 
However, our generative physics-embedded inverse analysis provides an approach to easily and rapidly conduct underground property characterization. 
Our approach generalizes several different inverse analysis approaches (e.g., PCGA and RegAE), and the accuracy of the final results depends on the choice of the generative model. 
Even if we need to discuss the more complicated problems in the future, the solution would be to improve the physical background understanding, which leads to applying an appropriate generative model to the related simulations. 
As a result, we only provide some fundamental insight into how to invert the underground geological structures by applying the inverse analysis method. 

Our analysis was performed on a machine with an Intel(R) Core(TM) i9-9960X CPU @ 3.10GHz with 32 threads and an NVIDIA RTX 2080 Ti GPU only for the VAE training.
Except for the VAE training for bimodal fields, all the other problems only need to prepare the three reference fields through the physics-embedded generative model, which does not require much generation time. 
However, the optimization is most time-consuming, and the time to perform the inverse analysis varies somewhat depending on the reference fields.
For example, for the Gaussian hydraulic conductivity problem, each epoch needs around 10 s to finish, while for the hydraulic fracture problem, the average time for each epoch is 10 m. 
As a result, the inverse analysis process time for all the problems mentioned in this study spans from 5 minutes to 1 day.
Furthermore, the gradient calculation dominates the total computational cost of the inverse analysis.
However, automatic differentiation efficiently improves the computation rate for the gradient calculation. 
Even if the reduction to $\mathbf{z}$ parameters from $\mathbf{p}$ interest properties makes the inverse analysis easy and fast, the application of automatic differentiation shows its extra benefit. 
The cost of computing a gradient with finite difference methods is $\sim$200 model runs, while the cost of computing a gradient including automatic differentiation is $\sim$2 model runs. 
Therefore, applying automatic differentiation helps speed up these computations by an additional factor of up to $\sim$100. 
In addition, the application of regularization allows for easy optimization. 
Combining these two features illustrates their vast potential for computational cost-saving for easy and efficient inverse analysis.

This study proposes the generative approach to include the physics-embedded generative model during inverse analysis.
The framework can efficiently characterize various underground properties inversion and demonstrate accurate and trustworthy prediction results.
Understanding the subsurface geological structure and properties helps in groundwater management and protection, oil and gas production estimation and optimization, and heterogeneous underground structure detection.
Our approach provides new avenues of support for achieving good performance for inverse analysis by including the physics-embedded generative model. 
In addition, with automatic differentiation, the optimization can be completed fast and efficiently.
Finally, we will explore more complicated and realistic geologic research problems by applying our proposed approach to expand its application in geologic properties inversion.

\section{Conclusion}\label{sec:conclusion}

We have presented the application of an inverse analysis approach with a physics-embedded generative model for underground geological properties characterization, which provides an efficient method of regularization and automatic differentiation.  
In this study, a novel method for inverse analysis is proposed, which generalizes different inverse analysis approaches, and we have tested the application of this approach for various problems. 
We used four physics-embedded generative models: one based on the principal components arising from the geostatistical structure of the parameter fields, another using a variational autoencoder that was trained on images of the parameter maps, a third that embeds the structure of a hydraulic fracturing well (including a relationship between fracture length and permeability), and a fourth that includes geologic layers with distinct P-wave velocities.
As a result, the physics-embedded inverse analysis provides accurate and consistent performance for various inverse problems. 
Using the physics-embedded generative model in combination with observational data enables to construction of a loss function that can be automatically differentiated.
Our approach is computationally efficient and obtains an excellent solution to the inverse problem by easing the regularization process and applying automatic differentiation.
In the future, different observational strategies need to be discussed to enhance the accuracy for more significantly complicated problems and deliver a high level of reliable inverse results based on an efficient observational plan.

\section*{Acknowledgement}
WH and DO acknowledge support from Los Alamos National Laboratory’s Laboratory Directed Research and Development Early Career Award (20200575ECR). 
SG acknowledges support from the United States Department of Energy through the Computational Science Graduate Fellowship (DOE CSGF) under grant number DE-SC0019323.

\section*{Data Availability}
A computer program automatically generated all the data used in this manuscript.
The code for generating the data, training the data, and performing the inverse analysis is available at https://github.com/OrchardLANL/Regularization-DP-paper.

\bibliographystyle{unsrt}
\bibliography{refs}

\end{document}